# Inapproximability After Uniqueness Phase Transition in Two-Spin Systems

Jin-Yi Cai[*]   Xi Chen[†]   Heng Guo[‡]   Pinyan Lu[§]

**Abstract**

A two-state spin system is specified by a matrix

$$\mathbf{A} = \begin{bmatrix} A_{0,0} & A_{0,1} \\ A_{1,0} & A_{1,1} \end{bmatrix} = \begin{bmatrix} \beta & 1 \\ 1 & \gamma \end{bmatrix}$$

where $\beta, \gamma \geq 0$. Given an input graph $G = (V, E)$, the partition function $Z_\mathbf{A}(G)$ of a system is defined as

$$Z_\mathbf{A}(G) = \sum_{\sigma: V \to \{0,1\}} \prod_{(u,v) \in E} A_{\sigma(u), \sigma(v)}. \tag{1}$$

We prove inapproximability results for the partition function in the region specified by the non-uniqueness condition from phase transition for the Gibbs measure. More specifically, assuming NP $\neq$ RP, for any fixed $\beta, \gamma$ in the unit square, there is no randomized polynomial-time algorithm that approximates $Z_\mathbf{A}(G)$ for $d$-regular graphs $G$ with relative error $\epsilon = 10^{-4}$, if $d = \Omega(\Delta(\beta, \gamma))$, where $\Delta(\beta, \gamma) > 1/(1 - \beta\gamma)$ is the uniqueness threshold. Up to a constant factor, this hardness result confirms the conjecture that the uniqueness phase transition coincides with the transition from computational tractability to intractability for $Z_\mathbf{A}(G)$. We also show a matching inapproximability result for a region of parameters $\beta, \gamma$ outside the unit square, and all our results generalize to partition functions with an external field.

---

[*]University of Wisconsin, Madison.
[†]Columbia University.
[‡]University of Wisconsin, Madison.
[§]Microsoft Research Asia.

# 1 Introduction

Spin systems are well studied in statistical physics and applied probability. We focus on two-state spin systems. An instance of a spin system is a graph $G = (V, E)$. A configuration $\sigma : V \to \{0, 1\}$ assigns to every vertex one of two states. The contributions of local interactions between adjacent vertices are quantified by

$$\mathbf{A} = \begin{bmatrix} A_{0,0} & A_{0,1} \\ A_{1,0} & A_{1,1} \end{bmatrix} = \begin{bmatrix} \beta & 1 \\ 1 & \gamma \end{bmatrix},$$

a $2 \times 2$ matrix with $\beta, \gamma \geq 0$. The partition function $Z_{\mathbf{A}}(G)$ of a system is defined as

$$Z_{\mathbf{A}}(G) = \sum_{\sigma : V \to \{0,1\}} \prod_{(u,v) \in E} A_{\sigma(u), \sigma(v)}, \tag{2}$$

and we use $\omega(G, \sigma) = \prod_{(u,v) \in E} A_{\sigma(u), \sigma(v)}$ to denote the weight of $\sigma$.

For a fixed $\mathbf{A}$, we are interested in the complexity of computing $Z_{\mathbf{A}}(G)$, where $G$ is given as an input. Many natural combinatorial counting problems can be formulated as two-state spin systems. For example, with $\beta = 0$ and $\gamma = 1$, $Z_{\mathbf{A}}(G)$ is exactly the number of independent sets (or vertex covers) of $G$. The definition of $Z_{\mathbf{A}}(G)$ in (2) can be generalized to larger $\mathbf{A}$, and the problem is also known as counting (weighted) graph homomorphisms [20, 16]. On the other hand, the Ising model is the special case where $\beta = \gamma$.

The *exact* complexity of computing $Z_{\mathbf{A}}(G)$ has been completely solved for any fixed symmetric matrix $\mathbf{A}$ [11, 3, 13, 6] and even for not necessarily symmetric $\mathbf{A}$ [8, 4, 2, 9, 7, 5] as part of the dichotomy theorems for general counting constraint satisfaction problems. When specialized to two-state spin systems, $Z_{\mathbf{A}}(G)$ is #P-hard to compute exactly, except for the two restricted settings of $\beta\gamma = 1$ or $\beta = \gamma = 0$, for which cases it is polynomial-time computable. Consequently, the study on two-state spin systems has focused on the approximation of $Z_{\mathbf{A}}(G)$, and this is the subject of the present paper.

Following standard definitions, a fully polynomial-time approximation scheme (FPTAS) for $Z_{\mathbf{A}}(G)$ is an algorithm that, given as input a graph $G$ and a parameter $\epsilon > 0$, outputs a number $Z$ that satisfies

$$(1 - \epsilon) \cdot Z_{\mathbf{A}}(G) \leq Z \leq (1 + \epsilon) \cdot Z_{\mathbf{A}}(G) \tag{3}$$

in time poly$(|G|, 1/\epsilon)$. A fully polynomial-time randomized approximation scheme (FPRAS) is then a randomized algorithm that, with probability $1 - \delta$, outputs a $Z$ satisfying (3) in time poly$(|G|, 1/\epsilon, \log(1/\delta))$.

For the Ising model, in a seminal paper [17] Jerrum and Sinclair gave an FPRAS for $Z_{\mathbf{A}}(G)$ when $\beta = \gamma > 1$. It was further extended to the entire region of $\beta\gamma > 1$ by Goldberg, Jerrum and Paterson [14]. A two-state spin system is called *ferromagnetic* if $\beta\gamma > 1$ and *anti-ferromagnetic* if $\beta\gamma < 1$. The approximability of $Z_{\mathbf{A}}(G)$ for anti-ferromagnetic systems is less well understood. Starting with counting independent sets in sparse graphs [10], the approximability of $Z_{\mathbf{A}}(\cdot)$ in bounded degree graphs is also widely studied. Significant progress has been made recently on the algorithmic side, and approximation algorithms for anti-ferromagnetic two-state spin systems have been developed in [24, 22, 19, 18], based on the technique of correlation decay introduced by Bandyopadhyay and Gamarnik [1] and Weitz [24]. Finally, a unified FPTAS was found [18] to approximate $Z_{\mathbf{A}}(\cdot)$ for all anti-ferromagnetic two-state spin systems of either bounded degree graphs or general graphs, when the system satisfies a *uniqueness condition*.

The uniqueness condition is named for, and closely related to, phase transitions that occur for the Gibbs measure. It depends on not only $\beta$ and $\gamma$ but also the degree of the underlying graph as well. Such phase transitions from statistical physics are believed to frequently coincide with the transitions of computational complexity from tractability to intractability. However, there are only very few examples where the conjectured link is rigorously proved. One notable example is for the hardcore gas model (or independent set with $\beta = 0$ and $\gamma = 1$), for which such a conjecture was rigorously proved (for almost all degree bounds) both for the algorithmic side [24] and for the hardness side [23, 12]. As discussed above [24, 22, 19, 18], for general anti-ferromagnetic two-state spin systems, the algorithmic part of the conjecture has recently been established. In this paper, we make substantial progress on the hardness part of the conjecture.

**Our Results.** For $\beta$ and $\gamma : 0 \leq \beta, \gamma \leq 1$ except at $(\beta, \gamma) = (0, 0)$ or $(1, 1)$, Goldberg, Jerrum and Paterson proved that the problem does not admit an FPRAS on general graphs (when there is no degree bound), unless NP = RP [14]. In their reduction, the degrees of the hard instances are unbounded. This is consistent with the uniqueness threshold conjecture. However, for any fixed $\beta, \gamma$ in the unit square, the uniqueness condition states that there exists a finite threshold degree $\Delta(\beta, \gamma)$ [22, 19, 18], which satisfies

$$\Delta(\beta, \gamma) > \frac{1 + \sqrt{\beta\gamma}}{1 - \sqrt{\beta\gamma}} = \frac{(1 + \sqrt{\beta\gamma})^2}{1 - \beta\gamma} \geq \frac{1}{1 - \beta\gamma}, \tag{4}$$

such that the system satisfies the uniqueness condition if the degree $d < \Delta(\beta, \gamma)$, and the non-uniqueness condition if $d \geq \Delta(\beta, \gamma)$. The paper [18] gives an FPTAS for graphs with degree bounded by $\Delta(\beta, \gamma)$. The conjectured coincidence of phase transition with hardness in complexity suggests that as soon as the degree of the input graph goes beyond $\Delta(\beta, \gamma)$, the problem becomes hard to approximate. Towards this direction we show that for any fixed $\beta, \gamma$ in the unit square, the problem does not have an FPRAS if the degree of the input graph is $\Omega(\Delta(\beta, \gamma))$, unless NP = RP. Our hardness also holds when restricted to input graphs that are regular. Formally, we prove the following theorem:

**Theorem 1.** *There exists a positive constant $h$ such that: Given any $\beta, \gamma : 0 \leq \beta, \gamma \leq 1$ with $(\beta, \gamma) \neq (0, 0)$, $(1, 1)$ and any integer $d \geq h/(1 - \beta\gamma)$, there is no randomized polynomial-time algorithm that approximates $Z_\mathbf{A}(G)$ in d-regular graphs $G$ with relative error $\epsilon = 10^{-4}$, unless NP = RP.*

Note the relation between our degree bound $h/(1 - \beta\gamma)$ and $\Delta(\beta, \gamma)$ from (4).

We also make progress on $(\beta, \gamma)$ outside the unit square. While the uniqueness condition is *monotone* inside the unit square, its behavior outside is significantly different. (See more discussions on this difference in Appendix A.) Without loss of generality, we consider the region defined by $\beta\gamma < 1$ with $0 < \beta < 1 < \gamma$. There is a uniqueness curve (see Figure 1), connecting the point $(1, 1)$ and the $\gamma$-axis. Above the curve, the system satisfies the uniqueness condition for any graph [19, 18]. Hence, hardness is only possible below the uniqueness curve. Furthermore, when $(\beta, \gamma)$ is outside the unit square but below this uniqueness curve, there is only a finite range of degrees $d$ for which the system does not satisfy the uniqueness condition. This makes it very challenging to prove a hardness result for them. Previously, the hardness was only obtained in [14] for a very tiny square $0 \leq \beta \leq \eta$ and $1 \leq \gamma \leq 1 + \eta$ where $\eta$ is roughly $10^{-7}$, near the point $(0, 1)$ corresponding to independent set or the hardcore gas model.

In this paper, we prove the following hardness result for $(\beta, \gamma)$ outside the unit square:

**Theorem 2.** *Given $\beta$ and $\gamma$ such that $0 < \beta < 1$, $\gamma > 1$ and $\beta\gamma < 1$, let*

$$\Delta' = \lceil -1/(\ln \beta + \ln \gamma) \rceil \quad \text{and} \quad \Delta^* = \lceil 1/\ln \gamma \rceil. \tag{5}$$

*When $\Delta^* \geq 8000\Delta'$, there is no randomized polynomial-time algorithm that approximates $Z_\mathbf{A}(G)$ in regular graphs $G$ of degree $\Delta^*$ with relative error $\epsilon = 10^{-4}$, unless NP = RP.*

The new hardness region is pictured in Figure 1.[1] Here the two white squares are the hardness regions acquired by Goldberg, Jerrum, and Paterson [14]. Beyond the uniqueness threshold, we know that FPTAS exists. Our hardness result, Theorem 2, applies to the region between the vertical line with $\gamma = 1$ and the curve to the left of the uniqueness threshold. Let us describe the new curve in more details. Again we focus on the region with $0 < \beta < 1 < \gamma$ and $\beta\gamma < 1$; There is a symmetric curve when $0 < \gamma < 1 < \beta$. Near the point $(1, 1)$, the condition imposed by Theorem 2 is almost linear. So the new curve is roughly a line with slope $-8000$ around $(1, 1)$. When the curve approaches the line of $\beta = 0$, $\Delta'$ becomes 1 and the condition requires $\gamma$ to be between 1 and roughly $1 + 1/8000$.

---
[1]The reader should be aware that, for illustration purposes, the picture is not drawn to actual scale.

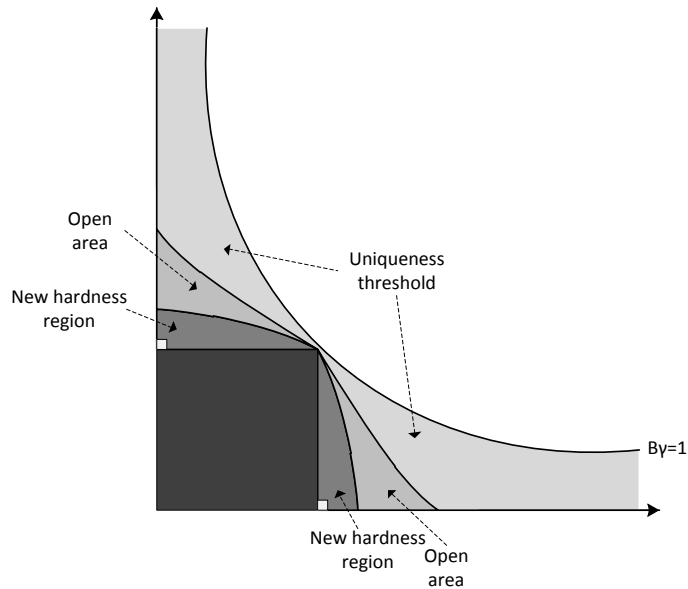

Figure 1: The new hardness region of Theorem 2.

Moreover, using a standard translation described in Appendix B we can generalize Theorem 1 and 2 to two-state spin systems with an external field. Formally, let $\mu \geq 0$, we have the following two corollaries for

$$Z_{\mathbf{A},\mu}(G) = \sum_{\sigma: V \to \{0,1\}} \mu^{\left|\{v \in V : \sigma(v) = 0\}\right|} \prod_{(u,v) \in E} A_{\sigma(u), \sigma(v)}.$$

**Corollary 1.** *There exists a constant $h$ such that, given any nonnegative $\beta, \gamma$ and $\mu$ with $\beta\gamma < 1$, and an integer $d$ satisfying that $\gamma \leq \mu^{\frac{1}{d}} \leq \frac{1}{\beta}$ and $d \geq \frac{h}{1 - \beta\gamma}$, there is no randomized polynomial-time algorithm that approximates $Z_{\mathbf{A},\mu}(G)$ in $d$-regular graphs $G$ with relative error $\epsilon = 10^{-4}$ unless $\mathrm{NP} = \mathrm{RP}$.*

**Corollary 2.** *Given any nonnegative $\beta, \gamma, \mu$ and an integer $d$ such that*

$$e^{\frac{1}{d}} \leq \gamma \cdot \mu^{-\frac{1}{d}} < e^{\frac{1}{d-1}} \qquad or \qquad e^{\frac{1}{d}} \leq \beta \cdot \mu^{\frac{1}{d}} < e^{\frac{1}{d-1}}$$

*if $d$ also satisfies $d \geq 8000 \lceil -1/(\ln \beta + \ln \gamma) \rceil$, then there is no randomized polynomial-time algorithm that approximates $Z_{\mathbf{A},\mu}(G)$ in $d$-regular graphs $G$ with relative error $\epsilon = 10^{-4}$ unless $\mathrm{NP} = \mathrm{RP}$.*

**Proof Outline.** In both Theorem 1 and 2, we use the phase transition that occurs in the non-uniqueness region to encode a hard-to-approximate problem. This approach has been used in previous hardness proofs for the hardcore gas model [10, 21, 23]. To this end, we reduce the approximation of E2LIN2 to the approximation of partition function in a two-state spin system. Here an instance of E2LIN2 consists of a set of variables $x_1, \ldots, x_n$ and a set of equations of the form $x_i + x_j = 0$ or $1$ over $\mathbb{Z}_2$. By [15], it is NP-hard to approximate the number of satisfiable equations for E2LIN2 within any constant factor better than $11/12$.

Given an E2LIN2 instance, we use a random bipartite regular graph to encode each variable $x_i$. Due to the phase transition and the fact that we are in a non-uniqueness region, each of these bipartite regular graphs would be in one of two types of configurations with high probability, if sampled proportional to its weight in the partition function. This can be used to establish a correspondence between the configurations of these bipartite graphs and the assignment of the boolean variables $x_1, \ldots, x_n$. Furthermore, we also add external connections between the bipartite graphs according to the set of equations in the E2LIN2 instance. They contribute exponentially to the total weight in the partition function, according to the total number of equations that an assignment satisfies. Thus, a sufficiently good approximation to the partition function can be used to decode approximately the maximum number of equations that an assignment can satisfy.

Our gadget is also randomly constructed. Then the probability should also be over the distribution of the gadgets. It is not hard to show that things work out beautifully if we simply substitute the expectation for the actual weight. But to make the proof rigorous, one must first obtain a sufficiently good concentration result. Such a result is unknown and could be very difficult to prove (assuming it is true), as it is already a tour-de-force in the special case for the hardcore gas model [21, 23, 12].

Instead we use a detour: (1) We prove a lower bound for the weights of two types of configurations we expect, guided by the phase transition; and (2) We prove that the total weight of other configurations is exponentially small compared to the lower bound with a probability exponentially close to 1. The way we establish the lower bound in (1) is similar to the approach by Dyer, Frieze, and Jerrum [10]. To prove (2), they [10] used the expectation and Markov's inequality. If we use the same approach, we could not get the hardness result for bounded degree graphs in the same order of the uniqueness bound. Instead, we use a new approach for (2).

In fact we first prove a high concentration result for an expander property of the gadgets we use. Then we show that the total weight of other configurations is exponentially small, given that the gadgets satisfy that property. This circumvented our inability to prove a complete concentration result. But we do prove some limited concentration results regarding the gadget, which let us prove hardness results for degrees in the right order conjectured according to the uniqueness threshold. It remains open whether one can use a refined version of this reduction along with the proof by Sly [23] to get the exact right bound. As discussed in Appendix A, this random regular graph quite closely follows the property of phase transition in infinite $d$-ary trees, when the parameter is below or beyond the uniqueness condition.

While the high-level idea of our proofs for both Theorem 1 and Theorem 2 are quite clear and similar, it remains a challenge to work out the estimation for all ranges of parameters and at the same time, make sure that the degree is in the same order of the uniqueness bound. To this end, technically we need to use very different approaches for Theorem 1 and Theorem 2. Even within Theorem 1 itself, we need to do the estimation differently for three different subcases.

## 2 Proof of the Main Theorems

From now on we will simply use $Z(G)$ to denote $Z_{\mathbf{A}}(G)$ or $Z_{\mathbf{A},\mu}(G)$ whenever it is clear from the context.

Given positive integers $N$ and $\Delta$, let $\mathcal{H}(N, \Delta)$ denote the following probability distribution of $\Delta$-regular bipartite graphs $H = (U \cup V, E)$ with bipartition $U, V$ and $|U| = |V| = N$: here $H$ is the union of $\Delta$ perfect matchings between $U$ and $V$ each selected independently and uniformly at random. (Because these perfect matchings are drawn independently, $H$ may have parallel edges.)

In both proofs of Theorem 1 and Theorem 2, we give a polynomial-time reduction from E2LIN2 to the approximation of $Z(G)$. An instance of E2LIN2 consists of $m$ equations over $\mathbb{Z}_2$ in $n$ variables $x_1, \ldots, x_n$. Each equation has exactly two variables and is of the form $x_i + x_j = b \in \{0, 1\}$. Without loss of generality we may always assume $m \geq n/2$; otherwise one of the variables does not appear in any equation. Given an assignment $S$ of the $n$ variables $x_1, \ldots, x_n$, we use $\theta(S)$ to denote the number of equations that $S$ satisfies and let $\theta^* = \max_S \theta(S)$. In [15] Håstad showed that it is NP-hard to estimate $\theta^*$ within any constant factor better than $11/12$.

Given an instance of E2LIN2 we construct a random $(\Delta + \Delta')$-regular graph $G$ as follows, with the two parameters $\Delta, \Delta'$ to be specified later. This construction is used in the proof of both Theorem 1 and 2:

**Construction of $G$ from an instance of E2LIN2.** For each variable $x_i$, $i \in [n]$, we let $U_i$ and $V_i$ denote two sets of $d_i m$ vertices each, where $d_i \geq 1$ denotes the number of equations in which $x_i$ appears (thus, $\sum_i d_i = 2m$). Moreover, $U_i$ and $V_i$ can be decomposed into

$$U_i = U_{i,1} \cup \cdots \cup U_{i,d_i} \quad \text{and} \quad V_i = V_{i,1} \cup \cdots \cup V_{i,d_i}$$

where each $U_{i,k}$ and $V_{i,k}$ contains exactly $m$ vertices. Now enumerate all the $m$ equations in the E2LIN2 instance one by one. For each of the $m$ equations do the following:

(1) Let $x_i + x_j = b \in \{0,1\}$ denote the current equation. Assume this is the $k$th time that $x_i$ appears in an equation, and the $\ell$th time that $x_j$ appears in an equation so far, where $k \in [d_i]$ and $\ell \in [d_j]$. Denote the $m$ vertices in $U_{i,k}$ by $\{u_1, \ldots, u_m\}$, vertices in $V_{i,k}$ by $\{v_1, \ldots, v_m\}$, vertices in $U_{j,\ell}$ by $\{u'_1, \ldots, u'_m\}$ and vertices in $V_{j,\ell}$ by $\{v'_1, \ldots, v'_m\}$. All these vertices have degree 0 at this moment. If $b = 0$, we add $\Delta'$ parallel edges between $(u_s, v'_s)$ and $(v_s, u'_s)$, for each $s \in [m]$; or if $b = 1$, we add $\Delta'$ parallel edges between $(u_s, u'_s)$ and $(v_s, v'_s)$, for each $s \in [m]$.

By the end of this step, every vertex has degree $\Delta'$. In the next step,

(2) For each $i \in [n]$, we add a bipartite graph $H_i = (U_i \cup V_i, E_i)$ drawn from $\mathcal{H}(d_i m, \Delta)$.

This finishes the construction, and we get a $(\Delta + \Delta')$-regular graph $G$ with $4m^2$ vertices.

We need the following notation. Given any assignment $\sigma$ from $V(G)$ to $\{0,1\}$, we let $U_i(\sigma)$ denote the number of vertices $u \in U_i$ with $\sigma(u) = 0$, and let $V_i(\sigma)$ denote the number of $v \in V_i$ with $\sigma(v) = 0$.

*Proof of Theorem 1.* Without loss of generality, we assume $0 \leq \beta \leq \gamma \leq 1$. We can also assume that $\beta > 0$ since the tight hardness to the exact uniqueness bound for $\beta = 0$ has been shown in [19] by generalizing the tight hardness result for the hardcore model [23, 12].

Given any assignment $S$ of the $n$ variables, we let $Z(G, S)$ denote the sum of $\omega(G, \sigma)$ over assignments $\sigma : V(G) \to \{0,1\}$ that satisfy for each $i \in [n]$,

$$U_i(\sigma) \leq V_i(\sigma) \text{ if } x_i = 0 \text{ in } S; \text{ or } U_i(\sigma) \geq V_i(\sigma) \text{ if } x_i = 1 \text{ in } S. \tag{6}$$

By definition, we have $Z(G, S) \leq Z(G) \leq \sum_S Z(G, S)$. We prove the following key lemma in Section 3:

**Lemma 1.** *There exists a positive constant $h$ with the following property: for any $\beta$ and $\gamma : 0 < \beta \leq \gamma \leq 1$ with $(\beta, \gamma) \neq (1,1)$ and for any $\Delta^* \geq h/(1 - \beta\gamma)$, there are $D > 1$, $C > 0$ and positive integers $\Delta$ and $\Delta'$ with $\Delta + \Delta' = \Delta^*$, that satisfy the following property: given any input instance of E2LIN2 with $n$ variables $x_1, \ldots, x_n$ and $m$ equations, except for probability $\leq \exp(-\Omega(m))$, the $\Delta^*$-regular graph $G$ constructed with parameters $\Delta$ and $\Delta'$ satisfies*

$$C^{m^2} \cdot D^{m\theta(S)} \leq Z(G, S) \leq C^{m^2} \cdot D^{m(\theta(S) + 0.03m)}, \quad \text{for any assignment $S$ of the $n$ variables.} \tag{7}$$

Given $\beta, \gamma$ and $\Delta^*$, we let $C, D, \Delta$ and $\Delta'$ denote the constants that satisfy the condition in Lemma 1, then given an input instance of E2LIN2, (7) holds with probability $1 - \exp(-\Omega(m))$.

Now assume (7) holds. We use $\theta^*$ to denote the maximum number of consistent equations and use $S^*$ to denote an assignment that satisfies $\theta^*$ equations. We also use $Y$ to denote an estimate of $Z = Z(G)$, where $|Y/Z - 1| \leq \epsilon = 10^{-4}$. From (7) and $Z(G, S^*) \leq Z(G) \leq \sum_S Z(G, S)$, we get

$$(1 + \epsilon) \cdot 2^n \cdot C^{m^2} \cdot D^{m(\theta^* + 0.03m)} \geq (1 + \epsilon) \cdot Z \geq Y \geq (1 - \epsilon) \cdot Z \geq (1 - \epsilon) \cdot C^{m^2} \cdot D^{m\theta^*} \tag{8}$$

Using $Y$, we set
$$Y' = \frac{\ln Y - \ln(1+\epsilon) - n \ln 2 - m^2 \ln C - 0.03 m^2 \ln D}{m \ln D}$$

and we get $Y' \le \theta^*$ since $\ln D > 0$. We finish the proof by showing that $Y' > (11/12) \cdot \theta^*$. By (8) we get

$$Y' \ge \theta^* - \frac{\ln(1+\epsilon) - \ln(1-\epsilon) + n \ln 2 + 0.03 m^2 \ln D}{m \ln D}$$

As $\theta^* \ge m/2$ and $m \ge n/2$, when $m$ is large enough, $Y' > (11/12) \cdot \theta^*$ and the theorem is proven. □

Next, we prove Theorem 2:

*Proof of Theorem 2.* For $\beta, \gamma$ with $0 < \beta < 1 < \gamma$ and $\beta\gamma < 1$, let $\Delta'$ and $\Delta^*$ be the two positive integers defined in (5) which satisfy $\Delta^* \ge 8000\Delta'$. We set $\Delta = \Delta^* - \Delta'$. Given any input instance of E2LIN2 with $n$ variables and $m$ equations, we use $G$ to denote the $\Delta^*$-regular graph constructed using $\Delta$ and $\Delta'$.

First of all, we show that to get a good approximation of $Z(G)$, with high probability it suffices to sum $\omega(G, \sigma)$ only over assignments $\sigma$ that satisfy the following condition:

$$\min\bigl(U_i(\sigma), V_i(\sigma)\bigr) \le \lambda d_i m, \quad \text{for all } i \in [n], \text{ where } \lambda = 9 \times 10^{-5}. \tag{9}$$

We use $\Sigma$ to denote the set of such assignments. Formally, we prove the following key lemma in Section 4:

**Lemma 2.** *Let $G$ be the graph constructed from an instance of E2LIN2 with $n$ variables $x_1, x_2, \ldots, x_n$ and $m$ equations, with parameters $\Delta$ and $\Delta'$. Then with probability $1 - \exp(-\Omega(m^{1/3}))$, it satisfies*

$$\sum_{\sigma \in \Sigma} \omega(G, \sigma) \le Z(G) \le \bigl(1 + o(1)\bigr) \cdot \sum_{\sigma \in \Sigma} \omega(G, \sigma). \tag{10}$$

Next, given an assignment $S$ over the $n$ variables we use $Z_\Sigma(G, S)$ to denote the sum of $\omega(G, \sigma)$ over all assignments $\sigma \in \Sigma$ that satisfy (6) for all $i \in [n]$. We prove the following lemma:

**Lemma 3.** *There are $C > 0$ and $D > 1$ satisfying the following property: given an instance of E2LIN2 with $n$ variables and $m$ equations, the $\Delta^*$-regular graph $G$ constructed with parameters $\Delta$ and $\Delta'$ satisfies*

$$C^{m^2} \cdot D^{m\theta(S)} \le Z_\Sigma(G, S) \le C^{m^2} \cdot D^{m\bigl(\theta(S) + 0.04m\bigr)}, \quad \text{for any assignment } S \text{ of the } n \text{ variables}. \tag{11}$$

*Proof.* Let $S$ be an assignment over the $n$ variables, and we use the same lower bound

$$Z_\Sigma(G, S) \ge Z^*(G, S) = C^{m^2} \cdot D^{m\theta(S)}$$

with $Z^*(G, S)$ defined in (15), $C$ and $D$ defined in (16). It is a lower bound for $Z_\Sigma(G, S)$ because every $\sigma$ that satisfies (14) is in $\Sigma$ by definition. Since $\beta\gamma < 1$, we have $C > 0$ and $D > 1$.

Now we give an upper bound for $Z_\Sigma(G, S)$. For each $\sigma$ in the sum $Z^*(G, S)$, we use $Q_\sigma$ to denote the following set of assignments $\sigma'$ in the sum of $Z_\Sigma(G, S)$: For each $i \in [n]$,

1. If $x_i = 0$ in $S$, then $\sigma'$ agrees with $\sigma$ on $V_i$ and $U_i(\sigma') \le \min\bigl\{V_i(\sigma), \lambda d_i m\bigr\}$, while $U_i(\sigma) = 0$;
2. If $x_i = 1$ in $S$, then $\sigma'$ agrees with $\sigma$ on $U_i$ and $V_i(\sigma') \le \min\bigl\{U_i(\sigma), \lambda d_i m\bigr\}$, while $V_i(\sigma) = 0$.

It is easy to show that $\{Q_\sigma\}$ is a partition of the assignments in $Z_\Sigma(G, S)$. Moreover, as in (21) we have:

$$|Q_\sigma| \le m^{2n} \cdot e^{2H(\lambda)M}$$

Every $\sigma'$ in $Q_\sigma$ has weight $\omega(G, \sigma') \le \omega(G, \sigma)$ because flipping a bit from 1 to 0 cannot increase its weight. Finally, we get the following bound for $Z_\Sigma(G, S)$:

$$Z_\Sigma(G, S) \le Z^*(G, S) \cdot m^{2n} \cdot e^{2H(\lambda)M}$$

To finish the proof, we plug in $\lambda = 9 \times 10^{-5}$ and $H(\lambda) < 0.000929$ to compare $2H(\lambda)$ with $0.04 \ln D$. Also recall the definition of $D$ in (16). From the assumption that

$$(\beta\gamma)^{\Delta'} \le 1/e \quad \text{and} \quad \gamma^{\Delta+\Delta'} < e\gamma < 1.001e$$

we get $\ln D > 0.04673$ and $0.04 \ln D > 2H(\lambda)$. (Here $\gamma < 1.001$ because by the condition of Theorem 1:

$$e > \gamma^{\Delta^*-1} \quad \text{and} \quad \Delta^* \ge 8000\Delta' \ge 8000$$

and thus, $\gamma < 1.001$.) This finishes the proof of the lemma. □

Let $\theta^* \ge m/2$ denote the maximum number of consistent equations, and let $S^*$ denote an assignment that satisfies $\theta^*$ equations. From these two lemmas, we have with high probability that

$$C^{m^2} \cdot D^{m\theta^*} \le Z_\Sigma(G, S^*) \le Z(G) \le \big(1 + o(1)\big) \cdot \sum_S Z_\Sigma(G, S) \le \big(1 + o(1)\big) \cdot 2^n \cdot C^{m^2} \cdot D^{m(\theta^*+0.04m)}$$

Theorem 2 then follows from the same argument used in the proof of Theorem 1. □

## 3 Proof of Lemma 1

Recall that $\beta$ and $\gamma$ satisfy $0 < \beta \le \gamma \le 1$ and $(\beta, \gamma) \ne (1, 1)$.

Given any instance of E2LIN2 with $n$ variables $x_1, \ldots, x_n$ and $m$ equations, the $(\Delta + \Delta')$-regular graph $G$ we construct consists of $\Delta$-regular bipartite graphs $H_i$, $i \in [n]$, and edges between them. For each $H_i$ in $G$, we use $U_i \cup V_i$ to denote its vertex set with $|U_i| = |V_i| = d_i m$. Recall that $\epsilon = 10^{-4}$. We say $A \subseteq U_i$ is *big* if $|A| \ge \epsilon|U_i|$, and $B \subseteq V_i$ is *big* if $|B| \ge \epsilon|V_i|$. We also use $E(H_i, A, B)$ to denote the number of edges between $A$ and $B$ in $H_i$. Let $K = 48/\epsilon^2 = 48 \times 10^8$. We prove the following lemma:

**Lemma 4.** *When $\Delta \ge K$, with probability $1 - \exp(-\Omega(m))$, the graph $G$ we construct satisfies*

$$E(H_i, A, B) \ge \frac{\Delta|A||B|}{4d_i m}, \quad \text{for all } i \in [n] \text{ and for all big } A \subseteq U_i \text{ and big } B \subseteq V_i. \tag{12}$$

*Proof.* We let $n$ denote $d_i m$ and let $d$ denote $\Delta$. Let $U, V$ be two sets of $n$ vertices. Let $H$ be a random $d$-regular bipartite graph generated by independently picking $d$ perfect matchings between $U$ and $V$ uniformly at random. We prove the following lemma. Lemma 4 then follows from Lemma 5 using the union bound.

**Lemma 5.** *Given any $A \subseteq U$ and $B \subseteq V$ with $|A| = an$ and $|B| = bn$, where $b \ge a \ge 10^{-4}$, we have*

$$\Pr\Big[\text{the number of edges between } A \text{ and } B \text{ in } H \le abdn/4\Big] \le 2^{-cdn}, \quad \text{where } c = 1/(16 \cdot 10^8).$$

*Proof.* The intuition is that $H$ is drawn from a distribution that is very close to $G(n, d)$.

We use $u_1, \ldots, u_{an}$ to denote the vertices in $A$. For each $k \in [d]$ and $i \in [an]$, we use $Y_{k,i}$ to denote the random $\{0, 1\}$-variable such that $Y_{k,i} = 1$ if $u_i$ is matched with a vertex in $B$ in the $k$th perfect matching; and is 0 otherwise. From this, we have

$$\Pr\Big[\text{the number of edges between } A \text{ and } B \text{ in } H \le abdn/4\Big] = \Pr\Bigg[\sum_{k\in[d],\,i\in[an]} Y_{k,i} \le abdn/4\Bigg].$$

However, the variables $Y_{k,1},\ldots,Y_{k,an}$ are clearly not independent.

To deal with this issue, we introduce the following independent random $\{0,1\}$-variables $X_{k,i}$, for every $k \in [d]$ and $i \in [an]$: $X_{k,i} = 1$ with probability $\rho_i = (bn - i + 1)/n$ and $X_{k,i} = 0$ with probability $1 - \rho_i$. It is easy to see that $Y_{k,i}$ dominates $X_{k,i}$: For all $k$ and $i$ and for all $y_{k,1},\ldots,y_{k,i-1} \in \{0,1\}$, we have

$$\Pr\Big[Y_{k,i} = 1 \,\Big|\, Y_{k,1} = y_{k,1}, \cdots, Y_{k,i-1} = y_{k,i-1}\Big] = \frac{bn - (y_{k,1} + \cdots + y_{k,i-1})}{n} \ge \rho_i = \Pr\Big[X_{k,i} = 1\Big].$$

Next we define, for each $k$, a sequence of $\{0,1\}$-random variables $Z_{k,i}$. They are highly correlated with the $X_{k,i}$'s but have the same joint distribution as $\{Y_{k,i} : i \in [an]\}$:

1. First, $Z_{k,1} = X_{k,1}$. Thus, $\Pr[Z_{k,1} = 1] = \rho_1 = b = \Pr[Y_{k,1} = 1]$.

2. For $Z_{k,2}$: If $X_{k,2} = 1$, then $Z_{k,2} = 1$; If $X_{k,2} = 0$ (which happens with probability $1 - \rho_2$), then we set $Z_{k,2}$ randomly as follows: Suppose $Z_{k,1} = z_{k,1} \in \{0,1\}$, then set $Z_{k,2} = 1$ with probability

$$w = \frac{(bn - z_{k,1})/n - \rho_2}{1 - \rho_2}$$

and set $Z_{k,2} = 0$ with probability $1 - w$. Note that we have

$$\Pr\Big[Z_{k,2} = 1 \,\Big|\, Z_{k,1} = z_{k,1}\Big] = \rho_2 + w(1 - \rho_2) = (bn - z_{k,1})/n.$$

Thus, $\{Z_{k,1}, Z_{k,2}\}$ has the same joint distribution as $\{Y_{k,1}, Y_{k,2}\}$. We also have $Z_{k,2} \ge X_{k,2}$.

3. For $Z_{k,i}$ from $i = 3$ to $an$: If $X_{k,i} = 1$, then we set $Z_{k,i} = 1$; If $X_{k,i} = 0$ (which happens with probability $1 - \rho_i$), then we set $Z_{k,i}$ randomly as follows: Suppose $Z_{k,j} = z_{k,j} \in \{0,1\}$ for each $j \in [i-1]$, then set $Z_{k,i} = 1$ with probability

$$w' = \frac{(bn - (z_{k,1} + \cdots + z_{k,i-1}))/n - \rho_i}{1 - \rho_i}$$

and set $Z_{k,i} = 0$ with probability $1 - w'$. Then note that

$$\Pr\Big[Z_{k,i} = 1 \,\Big|\, Z_{k,1} = z_{k,1}, \cdots, Z_{k,i-1} = z_{k,i-1}\Big] = \rho_i + w'(1 - \rho_i) = \Big(bn - (z_{k,1} + \cdots + z_{k,i-1})\Big)\Big/n$$

Thus, $\{Z_{k,1},\ldots,Z_{k,i}\}$ has the same joint distribution as $\{Y_{k,1},\ldots,Y_{k,i}\}$. We also have $Z_{k,i} \ge X_{k,i}$.

From the construction above it is clear that the joint distribution of $\{Z_{k,i} : i \in [an]\}$ is the same as that of $\{Y_{k,i} : i \in [an]\}$. Because $Z_{k,i} \ge X_{k,i}$ for all $i \in [an]$, we have

$$\Pr\Big[\sum Y_{k,i} \ge abdn/4\Big] = \Pr\Big[\sum Z_{k,i} \ge abdn/4\Big] \ge \Pr\Big[\sum X_{k,i} \ge abdn/4\Big]. \tag{13}$$

By (13), it now suffices to prove an upper bound for

$$\Pr\Big[\sum X_{k,i} \le abdn/4\Big].$$

We can now use the Chernoff bound. The expectation is $\mu = \sum_{k,i} \mathbf{E}[X_{k,i}] = d \cdot \sum_i \rho_i \geq abdn/2$. By the Chernoff bound (and setting $\delta = 1/2$), we have

$$\Pr\Big[\sum X_{k,i} \leq abdn/4\Big] \leq \Pr\Big[\sum X_{k,i} \leq (1-\delta)\mu\Big] \leq \exp(-\mu\delta^2/2).$$

The lemma then follows from

$$\exp\big(-\mu\delta^2/2\big) \leq \exp\left(-\frac{abdn}{2}\left(\frac{1}{2}\right)^2 \cdot \frac{1}{2}\right) \leq \exp\left(-\frac{10^{-8}dn}{2}\left(\frac{1}{2}\right)^2 \cdot \frac{1}{2}\right)$$

by setting $c$ to be $1/(16 \cdot 10^8)$. □

Using Lemma 5, we can then choose a large enough $d$ and apply the union bound on $A$ and $B$. □

To finish the proof of Lemma 1, we divide $(\beta, \gamma)$ into three cases. For each case we show there exists a large enough constant $h$ with the following property: for all $(\beta, \gamma)$ of this case, and for all $\Delta^* \geq h/(1-\beta\gamma)$, there are $C > 0$, $D > 1$ and positive integers $\Delta \geq K$ and $\Delta' \geq 1$ with $\Delta + \Delta' = \Delta^*$, such that (7) holds whenever $G$ satisfies (12). The lemma then follows by taking the maximum of the three $h$'s.

Let $L = 12/\epsilon^2 = K/4 = 12 \times 10^8$. In the rest of the proof, we assume $G$ satisfies (12) and let $M = m^2$.

## 3.1 Case 1: $0 < \beta < 1/2$ and $\beta \leq \gamma^L$

We set $h$ to be a large enough constant so that $h/(1-\beta\gamma) \geq 7(L+1)$. Given a $\Delta^* \geq h/(1-\beta\gamma)$, we then set $\Delta = \lfloor L\Delta^*/(L+1) \rfloor > K$ and $\Delta' = \lceil \Delta^*/(L+1) \rceil \geq 7$, where $\Delta + \Delta' = \Delta^*$ and $L\Delta' \geq \Delta \geq L(\Delta' - 1)$.

Let $S$ be an assignment over the $n$ variables $x_1, \ldots, x_n$, then we start with a lower bound $Z^*(G, S)$ for $Z(G, S)$. To this end, we consider the sum of $\omega(G, \sigma)$ over all assignments $\sigma$ that satisfy for each $i \in [n]$:

$$U_i(\sigma) = 0 \text{ if } x_i = 0 \text{ in } S; \text{ and } V_i(\sigma) = 0 \text{ otherwise}. \tag{14}$$

Denote this sum by $Z^*(G, S)$. It is clearly a lower bound for $Z(G, S)$, and is exactly equal to

$$\begin{aligned}
Z^*(G, S) &= \left(1 + 2\gamma^{\Delta+\Delta'} + \gamma^{2\Delta+2\Delta'}\right)^{m\theta(S)} \cdot \left(\beta^{\Delta'}\gamma^{\Delta'} + 2\gamma^{\Delta+\Delta'} + \gamma^{2\Delta+2\Delta'}\right)^{m(m-\theta(S))} \\
&= \left(\beta^{\Delta'}\gamma^{\Delta'} + 2\gamma^{\Delta+\Delta'} + \gamma^{2\Delta+2\Delta'}\right)^M \cdot \left(\frac{1 + 2\gamma^{\Delta+\Delta'} + \gamma^{2\Delta+2\Delta'}}{\beta^{\Delta'}\gamma^{\Delta'} + 2\gamma^{\Delta+\Delta'} + \gamma^{2\Delta+2\Delta'}}\right)^{m\theta(S)}.
\end{aligned} \tag{15}$$

Setting $C$ and $D$ appropriately, we have $Z^*(G, S) = C^M \cdot D^{m\theta(S)}$, where

$$C = \beta^{\Delta'}\gamma^{\Delta'} + 2\gamma^{\Delta+\Delta'} + \gamma^{2\Delta+2\Delta'} > 0 \text{ and } D = \frac{1 + 2\gamma^{\Delta+\Delta'} + \gamma^{2\Delta+2\Delta'}}{\beta^{\Delta'}\gamma^{\Delta'} + 2\gamma^{\Delta+\Delta'} + \gamma^{2\Delta+2\Delta'}} > 1 \tag{16}$$

since $(\beta, \gamma) \neq (0, 0), (1, 1)$. It is also easy to give a lower bound of $8/7$ for $D$ because the difference between the numerator and the denominator is $1 - \beta^{\Delta'}\gamma^{\Delta'} > 1/2$ as $\beta < 1/2$; and the denominator of $D$ is $< 7/2$.

Next, to give an upper bound for $Z(G, S)$, we consider the sum of $\omega(G, \sigma)$ over $\sigma$ that satisfies (6) and

$$U_i(\sigma) \leq \epsilon d_i m \text{ if } x_i = 0; \text{ and } V_i(\sigma) \leq \epsilon d_i m \text{ if } x_i = 1 \tag{17}$$

for every $i \in [n]$. We show that this sum is indeed a good approximation of $Z(G, S)$:

$$Z(G, S) \leq (1 + o(1)) \sum_{\sigma \text{ satisfies (6) and (17) for all } i} \omega(G, \sigma) \qquad (18)$$

To show (18) we randomly draw an assignment $\sigma$ from those appear in the sum $Z(G, S)$ with probability proportional to $\omega(G, \sigma)$, and it suffices to show that the probability that $\sigma$ satisfies (17) for all $i$ is $1 - o(1)$. This then follows from the following lemma and the union bound:

**Lemma 6.** *For any $i \in [n]$, the probability that $\sigma$ violates (17) is at most $\exp(-d_i m)$.*

*Proof.* Without loss of generality, we assume $x_i = 0$ in $S$.

Pick any partial assignment $\sigma'$ over all vertices of $G$ except those of $H_i$ which satisfies (6) for all $j \neq i$. To prove the lemma, it suffices to show that the sum of $\omega(G, \sigma)$ over all assignments $\sigma$ that are consistent with $\sigma'$ and satisfy (6) for $i$ but violate (17) at $i$ is exponentially smaller than $\omega(G, \sigma^*)$, where we use $\sigma^*$ to denote the unique assignment that is consistent with $\sigma'$ and satisfies $U_i(\sigma^*) = 0$ and $V_i(\sigma^*) = d_i m$.

To this end, we let $\omega(\sigma')$ denote the product of the edge weights in $\sigma'$ over all edges in $G$ except those have at least one vertex in $H_i$. Then it is easy to give the following lower bound for $\omega(G, \sigma^*)$:

$$\omega(G, \sigma^*) \geq \omega(\sigma') \cdot (\beta\gamma)^{\Delta' d_i m} \qquad (19)$$

On the other hand, for any assignment $\sigma$ that satisfies (6) (hence, $\sigma$ satisfies $V_i(\sigma) \geq U_i(\sigma)$, as we assumed $x_i = 0$), is consistent with $\sigma'$ but violates (17), we must have

$$\omega(G, \sigma) \leq \omega(\sigma') \cdot \beta^{\epsilon^2 \Delta d_i m / 4}. \qquad (20)$$

It follows from the assumption of (12) and $V_i(\sigma) \geq U_i(\sigma) \geq \epsilon d_i m$ as $x_i = 0$ and $\sigma$ violates (17).

Plugging in $\Delta$ and $\Delta'$, we have

$$\frac{\omega(G, \sigma^*)}{\omega(G, \sigma)} \geq \frac{\omega(\sigma') \cdot (\beta\gamma)^{\Delta' d_i m}}{\omega(\sigma') \cdot \beta^{\epsilon^2 \Delta d_i m / 4}} \geq \left(\frac{\beta^{2\Delta'}}{\beta^{3(\Delta'-1)}}\right)^{d_i m} = \left(\frac{1}{\beta^{\Delta'-3}}\right)^{d_i m} > 2^{3 d_i m}.$$

The lemma follows because the number of $\sigma$ that is consistent with $\sigma'$ but violates (17) is at most $2^{2 d_i m}$. $\square$

We continue with (18). For each $\sigma$ that satisfies (14), let $T_\sigma$ denote the following set of assignments $\sigma'$: (i) $\sigma'$ satisfies (17) for all $i$; and (ii) for each $i$, $\sigma'$ agrees with $\sigma$ over $V_i$ if $x_i = 0$; and agrees with $\sigma$ over $U_i$ if $x_i = 1$. It is clear that $\{T_\sigma\}$ is a partition of the assignments that satisfy (17) for all $i \in [n]$. It is easy to check that for any $\sigma$ that satisfies (14), we know exactly the cardinality of $|T_\sigma|$:

$$|T_\sigma| = \prod_{i \in [n]} \left(\sum_{j=0}^{\lfloor \epsilon d_i m \rfloor} \binom{d_i m}{j}\right) \leq \prod_{i \in [n]} \left((\lfloor \epsilon d_i m \rfloor + 1) \binom{d_i m}{\lfloor \epsilon d_i m \rfloor}\right) \leq m^{2n} \prod_{i \in [n]} e^{H(\epsilon) d_i m} = m^{2n} \cdot e^{2H(\epsilon) M} \qquad (21)$$

where $H(\epsilon) \approx 0.00102$. For any $\sigma' \in T_\sigma$, we also have

$$\omega(G, \sigma') \leq \omega(G, \sigma) \cdot \left(1/\gamma^{\Delta+\Delta'}\right)^{2\epsilon M}, \qquad (22)$$

because for any assignment, switching the value of a vertex from 1 to 0 can improve $\omega(G, \sigma')$ by at most a factor of $1/\gamma^{\Delta+\Delta'}$. Finally, by combining (18), (21) and (22) we get the following upper bound for $Z(G, S)$:

$$Z(G, S) \leq (1 + o(1)) \cdot m^{2n} \cdot e^{2H(\epsilon) M} \cdot Z^*(G, S) \cdot \left(1/\gamma^{\Delta+\Delta'}\right)^{2\epsilon M}. \qquad (23)$$

To finish the proof and show (7): $Z(G, S) < Z^*(G, S) \cdot D^{0.03M}$, we also need to compare $D$ with $1/\gamma^{\Delta+\Delta'}$. Since $\beta \leq \gamma^L$, we have $\beta^{\Delta'} \leq \gamma^{L\Delta'} \leq \gamma^\Delta$. It follows from the definition of $D$ that $D \geq 1/(4\gamma^{\Delta+\Delta'})$. Then (7) follows directly from (23) by plugging in $D > 8/7$ and $\epsilon = 10^{-4}$.

## 3.2  Case 2: $\beta \geq 1/2$ and $\beta \leq \gamma^L$

We set $h$, $\Delta$ and $\Delta'$ as follows. We pick $h$ to be a large enough constant so that for any $\Delta^* \geq h/(1 - \beta\gamma)$, $\Delta = \lfloor L\Delta^*/(L+1) \rfloor$ and $\Delta' = \lceil \Delta^*/(L+1) \rceil$ satisfy $\Delta' \geq 7$ and $(\beta\gamma)^{\Delta'} < 2^{-12}$. Since $\gamma \geq \beta$, we also have $\beta^{\Delta'} < 1/64$. By the definition of $\Delta$ and $\Delta'$, we have $L\Delta' \geq \Delta \geq L(\Delta' - 1)$.

The proof follows the same flow as that for Case 1. First of all, we use the same lower bound in (15):

$$Z(G, S) \geq Z^*(G, S) = C^M \cdot D^{m\theta(S)}$$

where $Z^*(G, S)$ is the sum of $\omega(G, \sigma)$ over $\sigma$ that satisfies (14). $C > 0$ and $D > 1$ are defined as in (16). It then follows from $(\beta\gamma)^{\Delta'} < 2^{-12}$ that $D > 4/(3 + 2^{-12}) \approx 4/3$.

We also use the same upper bound argument (18) for $Z(G, S)$:

$$Z(G, S) \leq (1 + o(1)) \sum_{\sigma \text{ that satisfies (17)}} \omega(G, \sigma). \tag{24}$$

To prove the same statement as in Lemma 6, pick an $i \in [n]$ and any partial assignment $\sigma'$ over all vertices of $G$ except those of $H_i$, satisfying (6) for all $j \neq i$. Without loss of generality assume $x_i = 0$ in $S$. We use $\sigma^*$ to denote the assignment that is consistent with $\sigma'$ and satisfies $U_i(\sigma^*) = 0$ and $V_i(\sigma^*) = d_i m$. We also use $\sigma$ to denote any assignment that is consistent with $\sigma'$, satisfies (6) in $H_i$, but violates (17). Then from (19) and (20), we have

$$\frac{\omega(G, \sigma^*)}{\omega(G, \sigma)} \geq \frac{(\beta\gamma)^{\Delta' d_i m}}{\beta^{\epsilon^2 \Delta d_i m/4}} \geq \left(\frac{\beta^{2\Delta'}}{\beta^{3(\Delta'-1)}}\right)^{d_i m} \geq \left(\frac{1}{\beta^{\Delta'/2}}\right)^{d_i m} > 2^{3d_i m}$$

The upper bound (24) then follows directly. To finish the proof we define $T_\sigma$ similarly for each assignment $\sigma$ that satisfies (17). Combining (24), (21) and (22), we obtain the same upper bound (23) for $Z(G, S)$. It also follows from $\beta \leq \gamma^L$ that $D > 1/(4\gamma^{\Delta+\Delta'})$. Then (7) is proven by plugging in $D \approx 4/3$ and $\epsilon = 10^{-4}$.

## 3.3  Case 3: $\beta > \gamma^L$

For this case, we need to use a different estimation for $Z(G, S)$.

We start by setting $\Delta'$ and $\Delta$. Let $h$ be a large enough constant such that for any $\Delta^* \geq h/(1 - \beta\gamma)$,

$$\Delta = \left\lceil \frac{L(L+1) \cdot \Delta^*}{L(L+1) + 1} \right\rceil \quad \text{and} \quad \Delta' = \left\lfloor \frac{\Delta^*}{L(L+1) + 1} \right\rfloor$$

satisfy $\Delta' \geq 1$ and $(\beta\gamma)^{\Delta'} < 1/4$. It follows from the definition that $\Delta^* = \Delta + \Delta'$ and $\Delta \geq L(L+1)\Delta'$.

Given an assignment $S$ over the $n$ variables $x_1, \ldots, x_n$, we use $\hat{\sigma}$ to denote the unique assignment with $U_i(\hat{\sigma}) = 0$ and $V_i(\hat{\sigma}) = d_i m$ when $x_i = 0$; $U_i(\hat{\sigma}) = d_i m$ and $V_i(\hat{\sigma}) = 0$ when $x_i = 1$, for all $i \in [n]$. Then

$$Z(G, S) > \omega(G, \hat{\sigma}) = (\beta\gamma)^{\Delta' m(m - \theta(S))} = \left((\beta\gamma)^{\Delta'}\right)^M \cdot \left(\frac{1}{(\beta\gamma)^{\Delta'}}\right)^{m\theta(S)} \tag{25}$$

Setting $C = (\beta\gamma)^{\Delta'}$ and $D = 1/C$, we get $Z(G, S) > C^M \cdot D^{m\theta(S)}$, with $D > 4$ and $C > 0$.

Next, to give an upper bound for $Z(G, S)$, we consider the sum of $\omega(G, \sigma)$ over $\sigma$ that satisfies

$$U_i(\sigma) \leq \epsilon \cdot d_i m \text{ and } V_i(\sigma) \geq (1 - \epsilon) \cdot d_i m, \quad \text{when } x_i = 0 \text{ in } S; \tag{26}$$

$$U_i(\sigma) \geq (1 - \epsilon) \cdot d_i m \text{ and } V_i(\sigma) \leq \epsilon \cdot d_i m, \quad \text{when } x_i = 1 \text{ in } S. \tag{27}$$

for every $i \in [n]$. We show that this sum is a good approximation of $Z(G, S)$:

$$Z(G, S) \leq (1 + o(1)) \sum_{\sigma \text{ that satisfies (26) and (27)}} \omega(G, \sigma) \tag{28}$$

To prove (28), we randomly draw a $\sigma$ from those appear in the sum $Z(G, S)$ with probability proportional to $\omega(G, \sigma)$, and show that the probability that $\sigma$ violates (26) or (27) is exponentially small.

For this purpose we prove the same statement as in Lemma 6. Pick any $i \in [n]$ and assume $x_i = 0$ in $S$ without loss of generality (the proof below handles (26), and (27) is by the same argument.) Let $\sigma'$ be any partial assignment over vertices of $G$ except those of $H_i$, and satisfying (6) for all $j \neq i$. We use $\sigma^*$ to denote the unique assignment that is consistent with $\sigma'$ and satisfies $U_i(\sigma^*) = 0$ and $V_i(\sigma^*) = d_i m$. We let $\sigma$ denote any assignment that is consistent with $\sigma'$ but violates (26) in $H_i$. Then we get

$$\frac{\omega(G, \sigma^*)}{\omega(G, \sigma)} \geq \frac{\omega(\sigma') \cdot (\beta\gamma)^{\Delta' d_i m}}{\omega(\sigma') \cdot \gamma^{\epsilon^2 \Delta d_i m / 4}} > \left(\frac{\gamma^{(L+1)\Delta'}}{\gamma^{3(L+1)\Delta'}}\right)^{d_i m} = \left(\frac{1}{\gamma^{2(L+1)\Delta'}}\right)^{d_i m} > 2^{4 d_i m}$$

Here the first inequality follows from (12) and the fact that, since $\sigma$ violates (26), either

$$V_i(\sigma) \geq U_i(\sigma) > \epsilon \cdot d_i m \quad \text{or} \quad U_i(\sigma) \leq V_i(\sigma) < (1 - \epsilon) \cdot d_i m \tag{29}$$

and the last inequality follows from $1/4 > (\beta\gamma)^{\Delta'} > \gamma^{(L+1)\Delta'}$. This proves (28).

Moreover, the number of $\sigma$ that satisfies (26) and (27) for all $i \in [n]$ can be bounded by $m^{4n} \cdot e^{4H(\epsilon)M}$, in the same way as (21). And for any $\sigma$ that satisfies both (26) and (27), we also have

$$\omega(G, \sigma) \leq \omega(G, \hat{\sigma}) \Big/ (\beta\gamma)^{2\epsilon \Delta' M} \tag{30}$$

This is because, to obtain $\sigma$ from $\hat{\sigma}$, each time we flip a vertex from 0 to 1, the weight increases by a factor of at most $1/\gamma^{\Delta'}$; each time we flip a vertex from 1 to 0, the weight increases by a factor of at most $1/\beta^{\Delta'}$.

Finally, combining (28), (29) and (30), we get

$$Z(G, S) \leq (1 + o(1)) \cdot m^{4n} \cdot e^{4H(\epsilon)M} \cdot \omega(G, \hat{\sigma}) \cdot D^{2\epsilon M}$$

Then (7) follows immediately by plugging in $D > 2$ and $\epsilon = 10^{-4}$. This finishes the proof of the lemma.

## 4   Proof of Lemma 2

Recall that $\beta$ and $\gamma$ satisfy $\beta, \gamma : 0 < \beta < 1 < \gamma$ and $\beta\gamma < 1$. Let $\Delta'$ and $\Delta^*$ be the two positive integers defined in Theorem 2, with $\Delta^* \geq 8000\Delta'$. From their definitions, we have $(\beta\gamma)^{\Delta'} \leq 1/e$ and $\gamma^{\Delta^*} \geq e$. Set $\Delta = \Delta^* - \Delta' \geq 7999\Delta' \geq 7999$. By the definition of $\Delta^*$, we have $e > \gamma^{\Delta^*-1} \geq \gamma^{\Delta}$ and thus, $\gamma < 1.001$.

Given an instance of E2LIN2 with $n$ variables $x_1, \ldots, x_n$ and $m$ equations, we use $G$ to denote the $\Delta^*$-regular graph constructed with parameters $\Delta$ and $\Delta'$, where $\Delta^* = \Delta + \Delta'$. We let $H_i$ denote the bipartite graph in $G$ that corresponds to $x_i$ and use $U_i \cup V_i$ to denote its vertices, with $|U_i| = |V_i| = d_i m$.

Before working on $G$ and $H_i$, we start by proving a property that a bipartite graph sampled from the distribution $\mathcal{H}(N, \Delta)$ satisfies with high probability. Let $H$ be a bipartite graph drawn from $\mathcal{H}(N, \Delta)$ for some $N \geq 1$ and $\Delta$ defined above, with $2N$ vertices $U \cup V$. We also use $\rho : U \cup V \to \{0, 1\}$ to denote an assignment and call it an $(a, b)$-assignment for some $a, b \in T_N$, where $T_N = \{0, 1/N, 2/N, \ldots, (N-1)/N, 1\}$ if $|u \in U : \rho(u) = 0| = aN$ and $|v \in V : \rho(v) = 0| = bN$. We also use $\mathcal{I}_N(a, b)$, where $a, b \in T_N$, to denote the set of all such $(a, b)$-assignments, and let

$$Z_{a,b}(H) = \sum_{\rho \in \mathcal{I}_N(a,b)} \omega(H, \rho) \cdot \gamma^{\Delta'(2-a-b)N} \tag{31}$$

with $\Delta'$ defined above. We are interested in the expectation of $Z_{a,b}(H)$ when $\min(a, b) \geq \lambda = 9 \times 10^{-5}$:

**Lemma 7.** *For large enough $N$ and $a, b \in T_N$ such that $\min(a, b) \geq \lambda$, we have*

$$\mathbf{E}_{H \leftarrow \mathcal{H}(N,\Delta)}\left[Z_{a,b}(H)\right] \leq \exp\left(1.21 \cdot N\right).$$

*Proof.* We recall the definition

$$Z_{a,b}(H) = \sum_{\rho \in \mathcal{I}_N(a,b)} \omega(H, \rho) \cdot \gamma^{\Delta'(2-a-b)N}.$$

We want to compute the expectation

$$E = \mathbf{E}_{H \leftarrow \mathcal{H}(N,\Delta)}\left[Z_{a,b}(H)\right]$$

Because the distribution $\mathcal{H}(N, \Delta)$ is symmetric for any $\rho \in \mathcal{I}_N(a, b)$, each term in the summation $Z_{a,b}(H)$ has exactly the same expectation. Thus, we have

$$E = \gamma^{\Delta'(2-a-b)N} \cdot \binom{N}{aN} \cdot \binom{N}{bN} \cdot \left( \sum_{k \in T_N, a+b-1 \leq k \leq \min(a,b)} \frac{\beta^{kN} \gamma^{(1-a-b+k)N} \binom{bN}{kN} \binom{(1-b)N}{(a-k)N}}{\binom{N}{aN}} \right)^\Delta.$$

Since we only care about the exponent of $E$ and the summation of $k$ is only over a linear number of terms we can replace the summation by maximum without changing the leading term of the exponent. Therefore we conclude that the coefficient of $N$ in the exponent of the expectation $E = \exp(\Psi(a, b)N)$ is of the following form, in which the function $H(x) = -x \ln x - (1 - x) \ln(1 - x)$: Let $c = 8000$, we have

$$\Psi(a, b)$$
$$= \max_k \left[ (2-a-b)\Delta' \ln \gamma + H(a) + H(b) + \Delta\left(k \ln \beta + (1-a-b+k)\ln \gamma + bH(\frac{k}{b}) + (1-b)H(\frac{a-k}{1-b}) - H(a)\right) \right]$$
$$= \max_k \left[ \Delta' \ln \gamma + (1-a-b)(\Delta + \Delta')\ln \gamma + H(a) + H(b) + \Delta\left(k \ln(\beta\gamma) + bH(\frac{k}{b}) + (1-b)H(\frac{a-k}{1-b}) - H(a)\right) \right]$$
$$\leq \max_k \left[ \frac{1}{c-1} + (1-a-b)\frac{c}{c-1} + H(a) + H(b) + (c-1)\left(k\Delta' \ln(\beta\gamma) + bH(\frac{k}{b}) + (1-b)H(\frac{a-k}{1-b}) - H(a)\right) \right]$$
$$\leq \max_k \left[ \frac{1}{c-1} + (1-a-b)\frac{c}{c-1} + H(a) + H(b) + (c-1)\left(-k + bH(\frac{k}{b}) + (1-b)H(\frac{a-k}{1-b}) - H(a)\right) \right]$$

For the last formula, we can use Mathematica to verify that its value is $< 1.21$ when $\min(a, b) \geq 9 \times 10^{-5}$. □

For now, by Lemma 7 we can impose the following condition on the graph $G$ constructed from the input instance of E2LIN2:

$$\text{For all } i \in [n] \text{ and all } a, b \in T_{d_i m} \text{ with } \min(a,b) \geq \lambda, \ Z_{a,b}(H_i) \leq \exp(1.22 \cdot d_i m). \tag{32}$$

Using Lemma 7, Markov's inequality and the union bound, it is easy to show that $G$ satisfies this condition with probability $1 - \exp(-\Omega(m))$. In the rest of the proof we show that $G$ satisfies (10) whenever it satisfies (32). Lemma 2 then follows immediately.

We assume that $G$ satisfies (32). To prove (10), we randomly sample an assignment $\sigma$ with probability proportional to $\omega(G, \sigma)$. (10) follows if we can prove that $\sigma$ satisfies (9) with probability $1 - o(1)$. For this purpose we need the following lemmas that give us properties that $\sigma$ satisfies with high probability. Given any set $L$ of vertices in $G$, we let $N_\sigma(L) = \{v \in L : \sigma(v) = 0\}$. Also recall the definition of $U_{i,k}$ and $V_{i,k}$ in the construction of $G$. Let $x_i$ be a variable and let $k \in [d_i]$, then

**Lemma 8.** *Let $\sigma$ be an assignment drawn according to its weight. For any $i \in [n]$ and any $k \in [d_i]$, except for probability $\exp(-\Omega(m^{1/3}))$, we have*

$$\left|N_\sigma(U_{i,k})\right| < \frac{1}{1+e} \cdot \left(1 + m^{-1/3}\right) \cdot |U_{i,k}|.$$

*Proof.* Pick any partial assignment $\sigma'$ over vertices of $G$ except those in $U_{i,k}$. Conditioned on $\sigma'$, it is easy to see that the values of vertices in $U_{i,k}$ are independent. Each vertex in $U_{i,k}$ has $\Delta + \Delta'$ neighbors, each of which contributes a vertex weight of either $\beta$ or $1$ if it is assigned $0$, and either $1$ or $\gamma$ if it is assigned $1$. Since $\gamma < 1/\beta$, the total weight for assignment $1$ is at least $\gamma^{\Delta+\Delta'} \geq e$ times the weight for assignment $0$. The lemma follows from the Chernoff bound. □

Given an assignment $\sigma$, we use $\sigma_i$ to denote its restriction over vertices in $H_i$ and $\sigma_{-i}$ to denote its partial assignment over vertices in $G$ except $H_i$. We let $M_{\sigma_{-i}}(U_i)$ denote the subset of $U_i$ whose unique neighbor outside of $H_i$ is assigned $1$. Using Lemma 8 and the union bound, we have

**Corollary 3.** *Let $\sigma$ be an assignment drawn according to its weight. Except for probability $\exp(-\Omega(m^{1/3}))$*

$$\left|M_{\sigma_{-i}}(U_i)\right| \geq \left(\frac{e}{1+e} - O(m^{-1/3})\right) \cdot |U_i|, \quad \text{for all } i \in [n]. \tag{33}$$

It is also clear that Lemma 8 and Corollary 3 also hold for $V_{i,k}$ and $V_i$, respectively, by symmetry. Now we are ready to prove Lemma 2. Let $\sigma = (\sigma_i, \sigma_{-i})$ be an assignment drawn from this distribution. Recall the definition of $\Sigma$ below (9). Then by Corollary 3 we have

$$\Pr\left[\sigma \notin \Sigma\right] \leq \exp(-\Omega(m^{1/3})) + \Pr\left[\sigma \notin \Sigma \ \Big|\ \sigma_{-i} \text{ satisfies (33) for both } U_i \text{ and } V_i \text{ and for all } i \in [n]\right] \tag{34}$$

To prove an upper bound for (34) we fix $\sigma_{-i}$ to be any partial assignment over the vertices of $G$ except those of $H_i$, which satisfies (33) for both $U_i$ and $V_i$. Then it suffices to prove that the sum of $\omega(G, \sigma)$ over all $\sigma \in \Sigma$ that are consistent with $\sigma_{-i}$, denoted by $Z_1$, is exponentially larger than the sum of $\omega(G, \sigma)$ over all $\sigma \notin \Sigma$ that are consistent with $\sigma_{-i}$, denoted by $Z_2$.

Let $\omega(\sigma_{-i})$ denote the product of the edge weights in $\sigma_{-i}$ over all edges that have no vertex in $H_i$. By the definition of $Z_{a,b}(H)$ in (31), we have

$$Z_2 \leq \omega(\sigma_{-i}) \sum_{a,b \in T_{d_i m}: a,b \geq \lambda} Z_{a,b}(H_i) \leq \omega(\sigma_{-i}) \cdot (d_i m)^2 \cdot \exp(1.22 \cdot d_i m), \tag{35}$$

where the second inequality follows from (32). To prove a lower bound for $Z_1$, we let

$$L = |M_{\sigma_{-i}}(U_i)| \quad \text{and} \quad R = |M_{\sigma_{-i}}(V_i)|.$$

Consider all assignments $\sigma$ that are consistent with $\sigma_{-i}$ and $U_i(\sigma) = 0$. This gives us

$$Z_1 \geq \omega(\sigma_{-i}) \cdot \gamma^{\Delta'L} \cdot (1 + \gamma^{\Delta+\Delta'})^R \cdot (\beta^{\Delta'} + \gamma^{\Delta})^{d_i m - R}.$$

By plugging in $\gamma^{\Delta+\Delta'} \geq e$, $\gamma^{\Delta} \geq e^{7999/8000}$, as well as the lower bound valid for $R$ in (33), we get

$$Z_1 \geq \omega(\sigma_{-i}) \cdot \exp(1.22897 \cdot d_i m),$$

and the lemma follows from (35).

## A  Uniqueness

Fix $\beta, \gamma > 0$ such that $\beta\gamma < 1$. We define the uniqueness condition and discuss some of its properties.

**Definition 1.** *Let $\hat{x}$ be the positive fixed point of the following function*

$$f(x) = \mu \left( \frac{\beta x + 1}{x + \gamma} \right)^d.$$

*We say that a tuple $(\beta, \gamma, \mu)$ exhibits uniqueness on d-regular graphs if*

$$\left| f'(\hat{x}) \right| = \frac{\mu d (1 - \beta\gamma)(\beta\hat{x} + 1)^{d-1}}{(\hat{x} + \gamma)^{d+1}} = \frac{d(1 - \beta\gamma)\hat{x}}{(\beta\hat{x} + 1)(\hat{x} + \gamma)} < 1.$$

This condition specifies whether the Gibbs measure of the system on the $(d+1)$-infinite regular tree is unique or not.

For a fixed tuple of parameters $(\beta, \gamma, \mu)$, if $(\beta, \gamma)$ lies inside of the unit square, i.e. $\beta, \gamma : 0 < \beta, \gamma \leq 1$, as $d$ increases, the uniqueness condition will eventually fail. In this region, the monotonicity property with respect to $d$ holds. That is, there exists some threshold $\Delta$ depending on $(\beta, \gamma, \mu)$ such that, for any $d \geq \Delta$, the uniqueness condition does not hold.

When one of $\beta, \gamma$ is larger than 1, to satisfy the uniqueness condition, the dependence between $(\beta, \gamma, \mu)$ and $d$ gets tricky. There exists a boundary such that, beyond it the uniqueness condition always holds. On the other hand, within the boundary, when $d$ increases, the uniqueness condition will eventually fail. But if $d$ gets even larger, the uniqueness condition will hold again.

## A.1 Random Regular Bipartite Graphs

A classical gadget to use is random regular bipartite graphs by combining $d$ perfect matchings. This is also the gadget we used in this paper. One intuitive reason why this gadget is good is because of its tree-like local structure. More formally, we can show that the expected behavior of this gadget undergoes a phase transition when the parameters of the system go across the uniqueness boundary. Let $Z_{a,b}$ be the expected weight summing over only subsets of size $an$ and $bn$ assigning 0 on each side respectively. If the system satisfies the uniqueness condition, then the maximum of $Z_{a,b}$ is achieved at the single point $(p^*, p^*)$; while if the system does not satisfy the uniqueness condition, $Z_{a,b}$ achieves its maximum at two points $(p^+, p^-)$, $(p^-, p^+)$, where $p^+ > p^-$. So the idea is to use these two maximum points to encode two states (a $\{0,1\}$ assignment of a variable, two parts of a cut, and so on), and reduce another problem to this.

The difficulty here is that the connection described above is only proved with respect to expectations. To make the reduction go through, we need to show that it holds with high probability, when we randomly choose a fixed gadget. Proving such a high concentration result, however, is not easy. For the special case of the hardcore model ($\beta = 0, \gamma = 1$) such a high concentration result was almost obtained after a sequence of work by a careful analysis of its second moment [21, 23, 12]. Proving such a result for general two-state spin systems seems very difficult.

Instead we observed that as $d$ increases, $p^+$ approaches 1 and $p^-$ approaches 0. Hence, as long as $d$ is large enough, we still manage to get some exponential gap between the weights of points near $(0, 1)$ or $(1, 0)$ and everywhere else that holds with high probability. We further argue that to achieve this gap, the degree $d$ differs with the desired uniqueness threshold by only a constant factor.

## A.2 The order of the threshold degree

Given the definition of the uniqueness condition, we are not able to give a closed form for the boundary $d$ of uniqueness and non-uniqueness in terms of $\beta, \gamma$ and $\mu$ in general. But some properties were known and we summarize them as follows. The following lemma is from [22] and [18]:

**Lemma 9.** *Assume $\beta, \gamma > 0$ and $\beta\gamma < 1$. If $\beta, \gamma$ and $d$ satisfy*

$$\sqrt{\beta\gamma} > \frac{d-1}{d+1} \quad \text{or equivalently,} \quad d < \frac{1+\sqrt{\beta\gamma}}{1-\sqrt{\beta\gamma}}$$

*the system always satisfies the uniqueness condition for any external field $\mu$.*

Because

$$\frac{1+\sqrt{\beta\gamma}}{1-\sqrt{\beta\gamma}} = \frac{(1+\sqrt{\beta\gamma})^2}{1-\beta\gamma} \geq \frac{1}{1-\beta\gamma},$$

the degree bound in Theorem 1 is tight up to a constant factor.

For $d > (1+\sqrt{\beta\gamma})/(1-\sqrt{\beta\gamma})$, we define

$$x_1(d) = \frac{-1 - \beta\gamma + d(1-\beta\gamma) - \sqrt{(-1-\beta\gamma+d(1-\beta\gamma))^2 - 4\beta\gamma}}{2\beta},$$

$$x_2(d) = \frac{-1 - \beta\gamma + d(1-\beta\gamma) + \sqrt{(-1-\beta\gamma+d(1-\beta\gamma))^2 - 4\beta\gamma}}{2\beta},$$

which are the two positive roots of the following equation

$$\frac{d(1-\beta\gamma)x}{(\beta x + 1)(x+\gamma)} = 1$$

The following lemma is from [18].

**Lemma 10.** *If $\gamma > \beta > 0$, $\beta\gamma < 1$ and $\sqrt{\beta\gamma} \le (d-1)/(d+1)$, then the system described by $(\beta, \gamma, \mu)$ exhibits uniqueness on d-regular graphs if and only if $\mu < \mu_1(d)$ or $\mu > \mu_2(d)$, where*

$$\mu_i(d) = x_i(d) \left(\frac{x_i(d) + \gamma}{\beta x_i(d) + 1}\right)^d, \quad \text{for } i = 1 \text{ and } 2.$$

When $0 \le \beta, \gamma < 1$ are treated as constants and $\mu$ approaches zero or infinite, we get the dependence of $d$ in terms of $\mu$ as follows:

- When $d$ is large, $x_1(d)$ is very small, $\mu_1(d)$ is in the order of $d\gamma^d$. To get non-uniqueness for very small $\mu$, we need that $\mu > \mu_1(d)$. This gives the bound $\log\mu/\log\gamma$.

- When $d$ is large, $x_2(d)$ is very large, $\mu_2(d)$ is in the order of $d/\beta^d$. To get non-uniqueness for very large $\mu$, we need that $\mu < \mu_2(d)$. This gives the bound $-\log\mu/\log\beta$.

Therefore, the dependence of $d$ in terms of $\mu$ in Corollary 1 is also tight up to a constant factor given that $0 \le \beta, \gamma < 1$ are treated as constants.

## B  Spin Systems with External Field

It is easy to verify that $Z_{\mathbf{A},\mu}(G)$ can be written as

$$Z_{\mathbf{A},\mu}(G) = \sum_{\sigma:V\to\{0,1\}} \mu^{s(\sigma)} \cdot \beta^{t_0(\sigma)} \cdot \gamma^{t_1(\sigma)}$$

where we use $s(\sigma)$ to denote the number of $v \in V$ with $\sigma(v) = 0$; $t_0(\sigma)$ to denote the number of $(u,v) \in E$ with $\sigma(u) = \sigma(v) = 0$; and $t_1(\sigma)$ to denote the number of $(u,v) \in E$ with $\sigma(u) = \sigma(v) = 1$.

Let $t_2(\sigma)$ denote the number of edges whose two ends are assigned different spin states in $\sigma$. For a regular graph of degree $d$, we have $d \cdot s(\sigma) = 2t_0(\sigma) + t_2(\sigma)$, and $t_0(\sigma) + t_1(\sigma) + t_2(\sigma) = |E|$. Thus we can get

$$s(\sigma) = \frac{2t_0(\sigma) + t_2(\sigma)}{d} = \frac{|E| + t_0(\sigma) - t_1(\sigma)}{d}$$

and rewrite $Z_{\mathbf{A},\mu}(G)$ as

$$Z_{\mathbf{A},\mu}(G) = \mu^{\frac{|E|}{d}} \sum_{\sigma:V\to\{0,1\}} \mu^{\frac{t_0(\sigma)-t_1(\sigma)}{d}} \cdot \beta^{t_0(\sigma)} \cdot \gamma^{t_1(\sigma)} = \mu^{\frac{|E|}{d}} \sum_{\sigma:V\to\{0,1\}} \left(\beta\mu^{\frac{1}{d}}\right)^{t_0(\sigma)} \cdot \left(\frac{\gamma}{\mu^{\frac{1}{d}}}\right)^{t_1(\sigma)}$$

The global factor $\mu^{\frac{|E|}{d}}$ can be computed in polynomial time and the summation part can be considered as the partition function on the same graph with the following new parameters

$$(\beta', \gamma', \mu') = \left(\beta\mu^{\frac{1}{d}}, \frac{\gamma}{\mu^{\frac{1}{d}}}, 1\right)$$

This is a two-state spin system without external field. We can apply Theorem 1 directly to get Corollary 1. To apply Theorem 2 we also need to further impose that the degree satisfies a certain relationship with the weight after transformation. This explains the conditions specified in Corollary 2.